
\magnification=1200
\baselineskip=12pt
\input Tables
\def\lsim{<\kern-2.5ex\lower0.85ex\hbox{$\sim$}\ }
\def\rsim{>\kern-2.5ex\lower0.85ex\hbox{$\sim$}\ }
\overfullrule=0pt
\def\LAMBDABAR {\hbox{$\lambda$\kern-0.52em\raise+0.45ex\hbox{--}\kern+0.2em}}
\def\ebar {\hbox{E\kern-0.6em\raise0.2ex\hbox{/}\kern+0.1em}}
\rightline{UR--1289\ \ \ \ \ \ \ \ }
\rightline{ER--40685--738}
\baselineskip=20pt
\centerline{\bf $\delta$--EXPANSION AND SELF--CONSISTENT CALCULATION}
\vskip 1cm
\centerline{Paulo F. Bedaque}
\centerline{and}
\centerline{Ashok Das}
\medskip
\centerline{Department of Physics and Astronomy}
\centerline{University of Rochester}
\centerline{Rochester, NY 14627, USA}
\vskip 2cm
\centerline{\underbar{Abstract}}

We compare results from $\delta$--expansion, in simple theories, with
self--consistent calculations as well as calculations involving the
principle of minimal sensitivity.  We show that the latter methods give
relatively more accurate results order by order.
\vfill\eject
\noindent {\bf I. \underbar{Introduction}:}

The $\delta$--expansion is an interesting idea [1,2] where one arranges the
perturbation series in powers of the interaction.  Thus, for example, one
can write the $\phi^4$--interaction in 3 + 1 dimension as
$$\eqalign{{\cal L}_I = - \lambda \phi^4 &= - \lambda M^4
\bigg( {\phi^4 \over M^4}\bigg) = \lim_{\delta \rightarrow 1} -
\lambda M^4 \bigg( {\phi^2 \over M^2}\bigg)^{1 + \delta}\cr
&= \lim_{\delta \rightarrow 1} - \lambda M^2 \phi^2 \bigg(
1 + \delta \ln \bigg( {\phi^2 \over M^2} \bigg) +
{1 \over 2!} \ \delta^2 \bigg( \ln \bigg( {\phi^2 \over M^2}\bigg)
\bigg)^2 + \dots \bigg) \cr}\eqno(1)$$
and develop a perturbation in powers of $\delta$.  Here $M$ is an arbitrary
mass scale introduced for convenience in obtaining a suitable expansion.
The $\delta$--expansion has several interesting consequences.  First,
expanding in powers of $\delta$ brings out nonperturbative effects in the
coupling constant $\lambda$.  This is quite easy to see from Eq. (1).
Namely, the expansion naturally introduces a mass term (quadratic term in
the field variables) where the mass is dependent on the coupling constant
and consequently, the propagator develops a nontrivial coupling constant
dependence which gives rise to the nonperturbative aspects in any
calculation.  This is, of course, very reminiscent of what happens in a
self-consistent calculation [3,4,5].  Furthermore, because the expansion
introdu
   ces
a mass term, it naturally avoids the infrared problems which plague
massless theories as well as theories at high temperature where masses can
be neglected.  Finally, it has ben shown, in simple models, that the
expansion in powers of $\delta$ leads to a series with radius of
convergence unity -- this is a much improved behavior over the naive
perturbation theory in the coupling constant which has zero radius of
convergence.

In spite of the many nice features, the $\delta$--expansion leaves some
questions unanswered.  For example, the mass scale $M$ introduced in the
theory is really an arbitrary scale and cannot be assigned any physical
meaning a priori.  The $\delta$-- expansion also appears to run into
difficulty when odd powers of the field operators are involved [6].
Furthermore, even though the radius of convergence of the
$\delta$--expansion series is unity, for a physical theory in 3 + 1
dimensions, the value of $\delta$ must be unity and higher for higher
dimensions.  The power series expansion in $\delta$, in this case, is known
to be unsatisfactory [2].  (The Pad\`e sum, however, is known to lead to
excellent results.)  Finally, since the $\delta$--expansion rearranges the
interaction Lagrangian into an infinite series, at any finite order it
would appear that such an expansion will not give meaningful results for
Greens functions even though it may lead to a very good approximation for
the generating functional.

As we indicated earlier, the $\delta$--expansion resembles self--consistent
calculational methods to some extent.  It was our goal, therefore, to
compare the self--consistent calculations with $\delta$--expansion
 to see if some of these criticisms can be answered by making such a
connection.  In the simple zero dimensional model which we study in this
paper, we find, on the other hand, that self--consistent calculations
consistently give better results order by order.  Our paper is organized as
follows.  In section II, we tabulate our results for the value of the
generating functional as well as the second moment up to fifth order in
 $\delta$--expansion.  In section III we carry out the same calculation in the
self--consistent method where the mass is determined self--consistently
(analogue of solving the gap equation) order by order up to fifth order.
In section IV we use the principle of minimal  sensitivity (P.M.S.) [7] to
deter
   mine the
value of the mass and carry out the calculation up to fifth order.  Our
conclusions are presented in section V.
\medskip
\noindent{\bf II.  \underbar{$\delta$--Expansion}:}

Let us consider the zero dimensional theory described by
$$Z (\delta) = \int {d \phi \over \sqrt{\pi}}\
e^{- \lambda (\phi^2)^{1 + \delta}}\eqno(2)$$
where $\lambda$ is the coupling constant of the theory.  This integral can
be exactly evaluated and has the value
$$Z_{\rm exact} (\delta) = e^{-E_{\rm exact}(\delta)} =
 {1 \over (\lambda)^{{1 \over 2 + 2 \delta}}}\ {2 \over \sqrt{\pi}}\
\Gamma \left( {3 + 2 \delta \over 2 + 2 \delta} \right) \, .\eqno(3)$$
The nonperturbative character of the exact generating functional is obvious
and let us note that for the $\phi^4$ and $\phi^6$ interactions
respectively, it has the values
$$\eqalign{Z_{\rm exact} (\delta = 1) &= 1.0228 \ {1 \over
 \lambda^{1/4}}\cr
Z_{\rm exact} (\delta = 2) &= 1.0457 \ {1 \over
 \lambda^{1/6}}\cr}\, .\eqno(4)$$
The Greens functions or the moments in this theory can also be exactly
calculated and let us note that the simplest moment has the value
$$G_{\rm 2\ exact}(\delta = 1) = \int
 {d \phi \over \sqrt{\pi}} \ \phi^2
e^{- \lambda \phi^4} = 0.3457\ {1 \over \lambda^{3/4}}\eqno(5)$$

We can, of course, calculate these quantities using the
$\delta$--expansion and we tabulate below the values up to fifth order in
the expansion.
\bigskip
\bigskip
\midinsert
\thicksize=.5pt
\thinsize=.5pt
\tablewidth=5.5in
\begintable
| 1st | 2nd | 3rd | 4th | 5th |Exact \crnorule
| Order | Order | Order | Order | Order | Value \cr
$Z(\delta = 1)$ | 0.9818 ${1 \over \lambda^{1/4}}$ | 1.1117 ${1 \over
\lambda^{1/4}}$|
0.8798 ${1 \over \lambda^{1/4}}$ |1.2140 ${1 \over \lambda^{1/4}}$ |0.7804 ${1
\
   over \lambda^{1/4}}$
 | 1.0228 ${1 \over \lambda^{1/4}}$\cr
$Z(\delta = 2)$ | 0.9635 ${1 \over \lambda^{1/6}}$ | 1.5045 ${1 \over
\lambda^{1/6}}$ | -0.3926  ${1 \over \lambda^{1/6}}$ |4.9529 ${1 \over
\lambda^{
   1/6}}$ |-8.9216 ${1 \over \lambda^{1/6}}$ |
1.0457 ${1 \over \lambda^{1/6}}$\cr
$G_2 (\delta = 1)$ |-0.0485 ${1 \over \lambda^{3/4}}$ |
1.8681 ${1 \over \lambda^{3/4}}$ | -1.6016 ${1 \over \lambda^{3/4}}$ |
4.3625 ${1 \over \lambda^{3/4}}$ | -5.55104 ${1 \over \lambda^{3/4}}$
| 0.3457 ${1 \over \lambda^{3/4}}$
\endtable
\vskip .1in
\endinsert\noindent
\noindent It is clear that order by order, the $\delta$--expansion results
do not quite agree with the exact results.  For the generating functional
it is known [2] that if one sums up the $\delta$--expansion series using the
Pad\` e approximation, then one obtains results quite close to the exact value
.  The alternating sign in the expression for $G_2  (\delta = 1)$ at
different orders similarly indicates that the series can be summed using
Pad\` e techniques and may give reasonable agreement with the exact result.
To obtain the Pade sum, however, one has to calculate the expansion
coefficients to higher orders.
\vfill\eject
\medskip
\noindent {\bf III. \underbar{Self--consistent Calculation}:}

Let us assume that the theory under consideration generates a dynamical
mass.  (In this case, in fact, since the integrals can be done exactly, we
know indeed that it does.)  In such a case, the dynamical mass can be
calculated self--consistently order by order by solving the gap equation
[3,4,5]
   .
One can then replace the original theory by a theory with a
self--consistently determined mass term and evaluate the generating
functional as a perturbation series in the coupling constant
$\lambda$.  At first sight, this may seem like a bad approximation since we
know that the exact values of the quantities we are interested in, have a
nonperturbative character.  But this, in fact, turns out to be a rather
good approximation since the self--consistent determination of the mass
develops a coupling constant dependence leading to the right
nonperturbative character of the results.

Technically, one way to implement the self--consistent method is to write [4]
$$Z(\delta) = \int {d \phi \over \sqrt{\pi}}
\ e^{-m^2 \phi^2 + \delta m^2 \phi^2 - \lambda (\phi^2)^{1 + \delta}}\, ,
\eqno(6)$$
where
$$\delta m^2 = m^2 (\delta) \eqno(7)$$
is to be determined order by order by requiring that the higher order
corrections to the 2 point 1PI function vanish.  Once this is determined, both
$\delta m^2 \phi^2$ and $- \lambda (\phi^2)^{1 + \delta}$ can be treated as
interactions and a perturbative expansion can be developed accurate to the
order in which the mass has been determined.  In solving the gap equation,
we find that the gap equation has a nontrivial solution only in alternate
orders and carrying out the calculations to fifth order, we obtain the
following results.
\vfill\eject
\bigskip
\bigskip
\midinsert
\thicksize=.5pt
\thinsize=.5pt
\tablewidth=5.5in
\begintable
| 1st | 3rd | 5th |Exact \crnorule
| Order | Order | Order |  Value \cr
${m (\delta = 1) \over \lambda^{1/4}}$
 | 1.31607  | 1.45875 |
1.58131 | \cr
$Z(\delta = 1)$ | 0.9498 ${1 \over \lambda^{1/4}}$ | 1.0089 ${1 \over
\lambda^{1/4}}$ | 1.0109 ${1 \over \lambda^{1/4}}$ |
1.0228 ${1 \over \lambda^{1/4}}$\cr
${m(\delta = 2) \over \lambda^{1/6}}$|1.4969 |1.8881 |2.1569 |\cr
$Z(\delta = 2)$| 0.8907 ${1 \over \lambda^{1/6}}$ |
0.9806 ${1 \over \lambda^{1/6}}$ | 1.0137 ${1 \over \lambda^{1/6}}$ |
1.0457 ${1 \over \lambda^{1/6}}$ \cr
$G_2(\delta = 1)$ | 0.2742 ${1 \over \lambda^{3/4}}$
| 0.3037 ${1 \over \lambda^{3/4}}$ | 0.3090 ${1 \over \lambda^{3/4}}$ |
0.3457 ${1 \over \lambda^{3/4}}$
\endtable
\vskip .1in
\endinsert\noindent
\noindent This table shows that the self--consistent calculations give a
very good approximation to the exact value for the generating functional as
well as the second moment order by order.  The perturbation series
is, in fact, quite well behaved with the results becoming closer to the
exact value at every higher order.
\medskip
\noindent {\bf IV. \underbar{Method of Minimum Sensitivity}:}

An alternate method for determining the mass is through the principle of
minimal sensitivity [7].  Once again, in this method we add and subtract a mass
term, treat one of the mass terms as well as the interaction term as
perturbation and develop a perturbation series.  The mass is then
determined order by order by requiring that the generating functional be
stationary with respect to small variations in the mass to that order,
namely,
$${dZ \over dm^2} = 0 \, .\eqno(8)$$
The drawback, in this method, is that the mass so determined does not
correspond to the physical mass.  However, this method leads to an
excellent agreement with the exact values -- even better than the agreement
for the self--consistent  method.  In this method also, we find nontrivial
solutions to the minimum condition only in alternate orders and the results
for our calculations are tabulated below.
\vfill\eject
\bigskip
\bigskip
\midinsert
\thicksize=.5pt
\thinsize=.5pt
\tablewidth=5.5in
\begintable
| 1st Order | 3rd Order | 5th Order |Exact \cr
${m (\delta = 1) \over \lambda^{1/4}}$
 | 1.25743  | 1.50619 |
1.67098 |\cr
$Z(\delta = 1)$ | 0.9573 ${1 \over \lambda^{1/4}}$ | 1.0109 ${1 \over
\lambda^{1/4}}$ | 1.0204 ${1 \over \lambda^{1/4}}$ |
1.0228 ${1 \over \lambda^{1/4}}$\cr
${m(\delta = 2) \over \lambda^{1/6}}$|1.4354 |1.8706 |2.1645 |\cr
$Z(\delta = 2)$| 0.8907 ${1 \over \lambda^{1/6}}$ |
0.9810 ${1 \over \lambda^{1/6}}$ | 1.0137 ${1 \over \lambda^{1/6}}$ |
1.0457 ${1 \over \lambda^{1/6}}$ \cr
$G_2(\delta = 1)$ | 0.3162 ${1 \over \lambda^{3/4}}$
| 0.3235 ${1 \over \lambda^{3/4}}$ | 0.3403 ${1 \over \lambda^{3/4}}$ |
0.3457 ${1 \over \lambda^{3/4}}$
\endtable
\vskip .1in
\endinsert\noindent
\noindent {\bf V. \underbar{Conclusion}:}

We have studied simple zero dimensional theories with
$\phi^4$ and $\phi^6$ interactions in $\delta$--expansion, in the
self--consistent
 method as well as through the principle of minimal sensitivity.
We find that order by order the self--consistent method as well as the
method of minimal sensitivity give better results than the
$\delta$--expansion both for the generating functional as well as the
second moment.  Even though we do not discuss it in this paper, it is quite
clear from our discussion that the self--consistent method at finite
temperature would lead to the exact result for the mass in the O(N) scalar
theory (unlike the $\delta$--
expansion) and would also avoid the problems of infrared divergence in the
massl
   ess
case just as the $\delta$--expansion does [8].

The calculations  presented in this paper were carried out using
MATHEMATICA.  One of us (P.B.) would like to thank CAPES for
partial financial support and Wen-Jui Huang for discussions
on PMS. This work was supported in part by U.S. Department of Energy
Grant No. DE--FG02--91ER40685.
\medskip
\noindent {\bf \underbar{References}}

\noindent
[1] C.M.Bender, K.A.Milton, M.Moshe, S.S.Pinsky and L.M.Simmons, Jr.,{\it
Phys.R
   ev.Lett.}{\bf 58}, 2615 (1987)

\noindent
[2] C.M.Bender, K.A.Milton, M.Moshe, S.S.Pinsky and L.M.Simmons, Jr., {\it
Phys.
   Rev.}{\bf D37}, 1472 (1988)

\noindent
[3] Y.Nambu and G.Jona--Lasinio, {\it Phys.Rev.} {\bf 122}, 345 (1961)

\noindent
[4] L.N.Chang and N.P.Chang, {\it Phys.Rev.}{\bf D29}, 312 (1984)

\noindent
[5] N.P.Chang and D.X.Li, {\it Phys.Rev.}{\bf D30},790 (1984)

\noindent
[6] C.M.Bender and K.A.Milton,{\it Phys.Rev.}{\bf D38}, 1310 (1988)

\noindent
[7] P.M.Stevenson, {\it Phys.Rev. }{\bf D23}, 2916 (1981)

\noindent
[8] C.M.Bender and A.Rebhan, {\it Phys.Rev.}{\bf D41}, 3269 (1990)
\end